\documentstyle{article}
\begin{document}
\LARGE

\begin{center}
\vspace*{0.3in} \bf How is a Black Hole Created from Nothing?
 \vspace*{0.6in}
\normalsize \large

\rm Zhong Chao Wu

Dept. of Physics

Zhejiang University of Technology

Hangzhou 310032, China

\vspace*{0.4in} \large \bf Abstract
\end{center}
\vspace*{.1in} \rm \normalsize \vspace*{0.1in}

Using the synchronous coordinates, the creation of a Schwarzschild black hole immersed in a de Sitter spacetime can be viewed as a coherent creation of a collection of timelike geodesics. The previously supposed conical singularities do not exist at the horizons of the constrained     instanton. Instead, the unavoidable irregularity is presented as a non-vanishing second fundamental form elsewhere at the quantum transition 3-surface. The same arguments can be applied to charged, topological or higher dimensional black hole cases.

 \vspace*{0.3in}
PACS number(s): 98.80.Hw; 04.50.+h; 04.62.+v


\vspace*{0.5in} \pagebreak

The primordial black hole problem has been extensively studied in the big bang, inflationary and quantum cosmological models. It is usually thought that a black hole is formed from a massive object through a gravitational collape scenario. However, the creation scenario of a black hole from nothing, i.e, from a seed instanton in quantum cosmology, is even much more dramatic. It is believed that there had existed an inflationary period in the very early universe, while spacetime was described by a de Sitter metric. Therefore, we are particularly interested in studying the quantum creation of a black hole with a de Sitter background.

In quantum cosmology, the wave function of the universe is defined as the path integral [1]
\begin{equation}
\Psi (h_{ij}, \phi_0) = \int d[g_{\mu\nu}]d[\phi] \exp( -I[g_{\mu\nu}, \phi]),
\end{equation}
where the 3-metric $h_{ij}$ and the matter fields $\phi_0$ on it are
the arguments of the wave function, and $g_{\mu\nu}, \phi$ are all closed
Euclidean 4-metrics and the matter fields on them with the variables of the wave function as the
only boundaries. $I$ is the Euclidean action
 \begin{equation}
 I = - \frac{1}{16\pi}\int_M\left  (R - 2\Lambda +L_m \right )- \frac{1}{8\pi}\oint_{\partial M} K,
 \end{equation}
where $R$ is the scalar curvature of the 4-metrics $M$, $K$ is the expansion rate of the boundary $\partial {M}$, $L_m$ is the Lagrangian of the matter fields $\phi$, and $\Lambda$ is the cosmological constant.
 
 The universe is created via a quantum transition from a seed instanton. It can be realized via an analytic continuation at the equator of the instanton. The relative creation probability of the universe, at the $WKB$ level,  is evaluated as
\begin{equation}
  P \approx \exp (- {I}).
 \end{equation}
 
 It can be derived from gravitational thermodynamics that the entropy of the system is equal to the negative of the action $I$ [2][3]. Therefore, the relative creation probability is the exponential of the entropy.
 
 The seed of a non-rotating black hole in the de Sitter spacetime background is a $S^2 $x$ S^2$ instanton. For a regular instanton, the Lorentzian spacetime created is the Nariai spacetime, which is interpreted as a pair of black holes with the creation probability of $\exp(-I) = \exp (2\pi\Lambda^{-1})$ [4]. In comparison, the creation probability of the de Sitter spacetime without a black hole is $\exp(-I) = \exp (3\pi\Lambda^{-1})$ [1].
 
 The Nariai black hole and the de Sitter spacetime are two cases of the de Sitter background with the greatest black hole and no black hole, respectively. Now we are interested in black hole creation with mass parameter $m$ interpolating these two extrema [5]. 
 
 It is known that an instanton should be an Euclidean regular solution of the Einstein field equation, therefore it should have a stationary action. However, for the generic black hole with the mass parameter $m$, the value of the Euclidean action for its Euclidean manifold, once made compact, depends on $m$. This means that the action is not stationary at least with respect to one degree of freedom, the parameter $m$ itself. Therefore, the associated instanton must be irregular [6]. 
 
 The metric of the Euclidean solution is
 \begin{equation}
 ds^2 = \triangle (r) d\tau^2 + \triangle^{-1} (r)dr^2 + r^2 d\Omega^2_2,
 \end{equation}
 where $d\Omega^2_2$ is the metric of the unit 2-sphere, and 
 \begin{equation}
 \triangle(r) = 1- \frac{2m}{r} - \frac{\Lambda r^2}{3}.
 \end{equation}

 The black hole horizon and the cosmological horizon are located at zeroes $r_l$ and $r_k$ of the rational expression $\triangle$. The surface gravity $\kappa_i$ of the horizon $r_i$ is $\mid d\triangle (r)/2dr\mid$ evaluated there. One can construct an instanton by identifying the $\tau$ coordinate with an arbitrary period $\beta$  on the two dimensional spacetime $(\tau, r)$, and then obtain the instanton between these two horizons. If one chooses $\beta = 2\pi \kappa^{-1}_i$, then the conical singularity at the horizon $i$ can be avoided. Of course, to regularize both conical singularities at the two horizons is impossible due to the reason mentioned above. The impossibility is represented by the fact that the surface gravities for two horizons are distinct. Even though, it can be shown as follows that the constructed manifold is of an action stationary under the condition that the 3-geometry of  the quantum transition equator  is given. The instanton is of topology $S^2$ x $S^2$, where one $S^2$ represents $d\Omega^2_2$, and the other $S^2$ is the distorted sphere in the spacetime $(\tau, r)$. Strictly speaking, the constrained instanton is the seed of the Schwarzschild-de Sitter spacetime  identified into one periodic cell in its Penrose-Carter diagram [5].

 Now one can evaluate the action of the manifold by recasting it into the canonical form [3][7]
 \begin{equation}
 I = I_l+I_k + \int_{M'}\left( \pi^{ij}\dot{h} _{ij}- NH_0 -N_iH^i \right)d^3x d\tau,
 \end{equation}
where $M^\prime$ is $M$ minus $M_l$ and $M_k$, here ${M_i}(i = l,k)$ denotes the small neighbourhood of horizon $r_i$ with a boundary of a constant coordinate $r$, and $N$ and $N_i$ are the lapse function and shift vector, $\pi^{ij}$ are the conjugate momenta of $h_{ij}$, $H_0$ and $H^{i}$ are the Einstein and momentum constraints, which vanish for all classical solutions of the Einstein equations, and dot denotes the derivative with respect to the imaginary time $\tau$, these associated terms vanish due to the $U(1)$ Killing time symmetry. Therefore, the action for $M^\prime$ equals zero.

On the other hand, the action $I_i$ can be expressed explicitly as
\begin{equation}
I_i = - \frac{1}{16\pi}\int_{M_i}\left  (R - 2\Lambda +L_m \right )- \frac{1}{8\pi}\oint_{\partial M_i} K.
\end{equation}
If there is a conical singularity at the horizon, its contribution to the action is reduced to the degenerate version of the second term, in addition to that from the boundary of $M_i$.

The Gauss-Bonnet theorem can be applied to the 2-dimensional $(\tau, r)$ section of $M_i$,
\begin{equation}
\frac{1}{4\pi}\int_{\hat{M}_i}\hat{R} + \frac{1}{2\pi}\oint_{\partial \hat{M}_i}\hat{K} + \frac{\delta_i}{2\pi} = \chi(i),
\end{equation}
where the hat notation represents the projection of those objects or quantities onto the 2-dimensional $(\tau, r)$ section, $\delta_i$ is the deficit angle of horizon $i$, and $\chi(i)$ is the Euler characteristic of $\hat{M}_i$. Since the expansion rate of the subspace $r^2 d \Omega^2_2$ goes to zero at the horizons. $K$ and $\hat{K}$ are equal.  Comparing eqs (7) and (8), as $M_i$ shrinks to the horizon, the action (7) approaches $-\chi(i)A_i/4$. where $A_i$ is the surface area of the horizon. It is noted that both the first terms in these equations vanish after shrinking.

Therefore, the entropy or the negative of the total action of the constrained instanton is [3]
\begin{equation}
S = - I = \frac{1}{4} \left ( \chi (l) A_l + \chi(k)A_k \right ).
\end{equation}
This is a quite universal formula. It is clear that the entropy of a black hole is originated from the topology of the instanton. 

The action is independent of the period parameter $\beta$. Indeed, from the derivation using canonical action form, it is obvious that the action is even independent of the way of gluing the south part and north part of the instanton. In the $(\tau, r)$ section one can join them along two arbitrary continuous curves connecting the two horizons with  discontinuities of the second fundamental forms, or even of the first fundamental form. The discontinuity of the second form is a jump of the three expansion, and the consequence of the discontinuity of the first form is not clear. It turns out that the crucial point is the existence of the two horizons.

From the above argument, it follows that there exist ambiguities in determining the quantum transition 3-surface, which are associated not only with arbitrariness of the period $\beta$. The question arises as to what is the true scenario of black hole creation? The goal here is to clarify this.

We are going to construct an alternative constrained instanton, which may be identified as the true instanton. The Schwarzschild-de Sitter black hole metric can be written in the synchronous coordinates as the following [8]:
\begin{equation}
ds^2 = - d\rho^2 + \frac{1}{\cos^2 \mu} \left ( \frac{\partial r(\rho,\mu)}{\partial \mu} \right )^2 d \mu^2 + r^2 (\rho, \mu) d\Omega^2_2.
\end{equation}

It shows that the classical evolution of the black hole is equivalent to a coherent motion of a collection of timelike geodesics with proper time $\rho$, labeled by $(\mu, \theta, \phi)$, in a potential hill described by $\triangle$, as shown by the following implicit transformation between the usual coordinates $(t,r)$ and $(\rho, \mu)$,
\begin{equation}
\rho  = \int_{r_0}^r \frac{dr}{[E^2 - \triangle]^{1/2}},    \;\;\;\; (E^2 = \cos^2 \mu),
\end{equation}
\begin{equation}
t = \int_{r_0}^r \frac{Edr}{[E^2 - \triangle]^{1/2}\triangle},
\end{equation}
where $r_0$ is an arbitrary constant,  which is associated with the gauge freedom of the synchronous coordinate form of the metric. However, it  will be specified for our discussion later.

It is noted that although  metric (10) appears to be time dependent with respect to $\rho$, it is time independent with respect to the Killing time $t$. The time $\rho$ is used for the convenience of discussing  the geodesics only.

Our alternative constrained instanton can be obtained by going to the regime where  $(E^2 - \triangle)\leq 0 $ and $(r_l \leq r \leq r_k)$, so the coherent motion occurs in the potential well $-\triangle$ in the imaginary synchronous time  and the spacetime is Euclidean. For future convenience, we shall use two complex time coordinates in the following discussion: $t + i\tau$ and $\rho + i\sigma$, so that the left hand sides of (11) and (12) should be replaced by  $\rho + i\sigma$ and $t + i\tau$ respectively. The quantum transition would occur when $ E^2 = \triangle$. Here the time derivative of $r$ vanishes, no matter which time is referred: $t, \tau, \rho$ or $\sigma$. When $(E^2 - \triangle)\geq 0 $ the coherent motion occurs in the real time and the spacetime becomes Lorentzian. The parameter $\mu$ of the geodesic is determined by the $r_b(\mu)$ value at the moment it is subject to the quantum transition, which is $\cos^2\mu = \triangle (r_b)$. For example, considering the geodesics passing through the horizons, one has $\cos \mu = 0$, or $\mu = \pi/2$.

The bottom of the potential well is located at
\begin{equation}
r_c= 3^{1/3}m^{1/3}\Lambda^{-1/3},
\end{equation}
and we define $\cos \mu_c \equiv  (1- 3^{2/3}m^{2/3}\Lambda^{1/3})^{1/2}, \left (\frac{\pi}{2}> \mu_c > 0 \right)$ and for the $S^1$ equator of the quantum transition  $(r_l \rightarrow r_c \rightarrow r_k \rightarrow r_c \rightarrow r_l)$ of $r$ values  correspond to $(\pi/2 \rightarrow \mu_c \rightarrow \pi/2 \rightarrow \mu_c \rightarrow \pi/2)$ of $\mu$ values.

We consider the bottom of the potential well $r = r_c$ as the south pole of the $(\tau, r)$ space, and specify $r_0= r_c$ for equations (11) and (12). The total expansion rate around  the south pole is
\[
\oint \frac{\partial g_{\mu\mu}^{1/2}}{\partial \sigma}d\mu 
= 4 \int_{\mu_c}^{\pi/2}\frac{1}{\cos \mu}\frac{\partial^2 r}{\partial \sigma \partial \mu} d \mu 
= 4 \int_{\mu_c}^{\pi/2}\frac{1}{\cos \mu}\frac{\partial^2 r}{\partial \mu \partial \sigma} d \mu
= 4 \int_{\mu_c}^{\pi/2}\frac{1}{\cos \mu}\frac{\partial [\triangle - E^2]^{1/2}}{\partial \mu}d\mu
\]
\begin{equation}
 = 4\int_{\mu_c}^{\pi/2}[\cos^2 \mu_c - \cos^2\mu]^{-1/2}\sin\mu d\mu  = 2\pi.
\end{equation}

This result implies that the south pole is regular, therefore by setting $r_0 = r_c$, one obtains an seed constrained instanton. If one chooses $r_0 \neq r_c$ as the south pole, then some irregularity must occur there, so (14) is not valid, this means that one does not get a seed instanton. Once we set $r_0 = r_c$, it follows from (11) and (12) that $t = \tau = \rho = \sigma =0 $ at the south pole.

A geodesic with given parameter $\mu$ can be thought of as a particle moving in the potential well, with $\sigma$ referred to as the Newtonian time. The time it takes from the bottom $r_c$ to the highest point $r_b(\mu)$ in the imaginary time is 
\begin{equation}
\sigma_b(\mu)  = \left | \int_{r_c}^{r_b(\mu)}\frac{dr}{ [\triangle - \cos^2\mu]^{1/2}}\right |   
\end{equation}
and
\begin{equation}
\tau_b(\mu) =\left | \int_{r_c}^{r_b(\mu)} \frac{\cos \mu dr}{[\triangle -\cos^2\mu]^{1/2}\triangle}\right |.
\end{equation}

 Since the well does not take the form of that of a linear oscillator, the time spent to reach the highest point depends on the parameter $\mu$. This fact implies that the quantum transitions do not occur simultaneously in the imaginary synchronous time $\sigma$. The same argument is valid in the imaginary time $\tau$. One can consider the role of $\triangle$ in (16) as a redshift effect of the gravitational field. In general, the second fundamental form at the boundary of the south part of the instanton does not vanish, except for those which happen to be located at the two horizons. In contrast, in the earlier literature, the irregularities are supposed to be concentrated as conical singularities at the horizons [5].

In the Lorentzian regime, on the other hand, the creation of all geodesics does occur simultaneously at $t=0$ in the real time. At the horizons $\tau_b$ is identified as one quarter of the inverse temperature there. Its significance elsewhere at the quantum transition 3-surface is not clear. How the fact that the quantum transitions for all geodesics do not occur simultaneously in the imaginary time $\tau$ is related to any observable effect is not known, at least for the time being. The above discussion suggests that describing a spacetime by complex coordinates may reveal richer phenomena in gravitational physics.

To recast the de Sitter spacetime creation into the synchronous coordinates is quite instructive. The metric is
\begin{equation}
ds^2 = - d\rho^2 + \frac{3}{\Lambda} \cosh^2 \left (\sqrt{\frac{\Lambda}{3}} \rho \right )d\mu^2 + 
 \frac{3}{\Lambda} \cosh^2 \left (\sqrt{\frac{\Lambda}{3}} \rho \right )\sin^2 \mu d\Omega^2_2.
\end{equation}
And the values of both (15) and (16) are $\sqrt{3}\pi/2\sqrt{\Lambda}$, which are independent of parameter $\mu$, since the potential $-\triangle = -1 +\frac{\Lambda r^2}{3}$ is of a linear oscillator form, whose period is independent of the amplitude. The instanton is formed by joining two standard hemispheres of $S^4$, as has been known for a long time [1]. The fact that integration value of (16) is independent of $\mu$ is well expected in physics, while quite surprising in mathematics, given the fact that the integrand in (16) is a rather complex function of $E$ or $\mu$.

Our discussion can be applied to any $d$-dimensional solution of the Einstein equation with metric form (4), in which $\triangle (r)$ becomes a more general function with two zeroes as black hole and cosmological horizons, and $\Omega^2_2$ is replaced by a $d-2$-dimensional homogeneous and isotropic space $\Omega^2_{d-2}$. For 4-dimensional topological black hole cases, the metrics take  form (4), with $\triangle$ and $\Omega^2_2$ replaced by
\begin{equation}
\triangle = p - \frac{2m}{r} - \frac{\Lambda r^2}{3}
\end{equation}
and 
\[
p=1,\;\;\;\;\;\; d\Omega^2_2 = d\theta^2 + \sin^2 \theta d\phi^2,
\]
\[
p=0,\;\;\;\;\;\;\;\;\;\;\; d\Omega^2_2 = d\theta^2 +  \theta^2 d\phi^2,
\]
\begin{equation}
p=-1,\;\; d\Omega^2_2 = d\theta^2 + \sinh^2 \theta d\phi^2,
\end{equation}
where the 2-spaces $\Omega^2_2$ are made compact topologically for flat $(p =0)$ and hyperbolic $(p= -1)$ cases.

The argument is also valid for charged black holes, then $\triangle$  of metrics (4) is replaced by
\begin{equation}
\triangle = p - \frac{2m}{r}  +\frac{Q^2}{r^2} - \frac{\Lambda r^2}{3},
\end{equation}
where $Q$ is the electromagnetic charge. If the charge is electric, a Legendre transform for the action of the instanton is required to  recover the duality between the magnetic and electric cases. It is noted that in the canonical form of the action, there is a Gauss constraint term, in addition to the momentum and Einstein constraints [3][9].

To generalize our argument to higher dimensional black holes with metric form (4) is straightforward, where $\Omega^2_2$ is replaced by $\Omega^2_{d-2}$ and
\begin{equation}
\triangle = p - \frac{\eta}{r^{d-3}} - \frac{r^2}{l^2},
\end{equation}
where $d$ is the dimensionality,  $\eta$ is proportional to the mass and $l$ is the radius of curvature of the de Sitter background [10].

In fact, our arguments can also be applied to the $BTZ$ or Lovelock black hole cases.

\vspace*{0.3in} \rm

\bf References:

\vspace*{0.1in} \rm

1. S.W. Hawking, in \it Astrophysical Cosmology\rm, Pontificial
Academial Scientiarum Scripta Varia \underline{48}, 563 (1982), eds.
H.A. Bruck, G.V. Coyne, and M.S. Longair. J.B. Hartle, and S.W. Hawking, \it Phys. Rev. \bf D\rm
\underline{28}, 2960 (1983).

2. G.W. Gibbons, and S.W. Hawking,  \it Phys. Rev. \bf D\rm
\underline{15}, 2738 (1977). G.W. Gibbons, and S.W. Hawking, \it
Phys. Rev. \bf D\rm \underline{15}, 2752 (1977). 

3. Z.C. Wu, \it Gen. Rel. Grav. \rm\underline{32}, 1823 (2000).

4. R. Bousso and S.W. Hawking, \it Phys. Rev. \bf D\rm \underline{52}, 5659 (1995). 

5. Z.C. Wu, \it Int. J. Mod. Phys. \bf D\rm\underline{6}, 199
(1997). R. Bousso and S.W. Hawking, \it Phys.Rev. \bf D\rm
\underline{59}, 103501 (1999); Erratum-ibid. \bf D\rm
\underline{60}, 109903 (1999).

6. Z.C. Wu, \it Gen. Rel. Grav. \rm\underline{30}, 1639 (1998). 

7. 	C. W. Misner, K. S. Thorne and J. A. Wheeler, \it Gravitation, \rm (W.H. Freeman, San Francesco, 1973); C. Teitelboim, \it Phys. Rev. \bf D\rm \underline{51}, 4315 (1995).

8. Z.C. Wu, in \it Proceeding of the Fourth Marcel Grossman
Meeting \rm edited by R. Ruffini (North Holland, Amsterdam, 1986).

9.  S.W. Hawking and S.F. Ross, \it Phys. Rev. \bf
D\rm\underline{52}, 5865 (1995). R.B. Mann and S.F. Ross, \it Phys.
Rev. \bf D\rm\underline{52}, 2254 (1995).

10. Oscar J.C. Dias and Jose P.S. Lemos, \it Phys. Rev. \bf
D\rm\underline{70}, 124023 (2004). Rob Myers, in \it Black Holes in Higher Dimensions \rm edited by Gary T. Horowitz (Cambridge University Press, 2017).

\end{document}